\documentclass[aps,a4paper,twocolumn,floatfix,showkeys,
groupaddress]{revtex4}

\usepackage{bm,amsmath,amssymb,amsfonts}

\usepackage[dvips]{graphicx}
\usepackage[colorlinks=true,citecolor=blue,urlcolor=red,filecolor=green]{hyperref}
\usepackage{color}
\definecolor{mycolor}{rgb}{0.30,0.50,0.20}
\begin{document}

\title{\textcolor{mycolor}
{Controlled imprisonment of wave packet and flat bands in a fractal geometry}}  

\author{Atanu Nandy}
\email{atanunandy1989@gmail.com}
\affiliation{Department of Physics, Kulti College, Kulti, Paschim Bardhaman,
West Bengal-713 343, India}


\begin{abstract}
The explicit construction of non-dispersive flat band modes and the tunability of has been reported for a hierarchical 3-simplex fractal geometry. A single band tight-binding Hamiltonian defined for the deterministic self-similar non-translationally invariant network can give rise to a countably infinity of such self localized eigenstates for which the wave packet gets trapped inside a characteristic cluster of atomic sites. An analytical prescription to detect those dispersionless states has been demonstrated elaborately. 
The states are localized over clusters of increasing sizes, displaying the existence of a multitude
of localization areas. The onset of localization can, in principle, be `delayed' in space by an appropriate
choice of the energy of the electron. 
The tunability of those states leads to the controlled decay of wave function envelope. The impact of perturbation on the bound states has also been discussed. The analogous wave guide model has also been discussed.
\end{abstract}
\maketitle
\section{Introduction}
\label{intro}
Flat band (FB)~\cite{suther} networks with the presence of one or  more diffraction-free momentum independent flat energy band(s) have grabbed considerable attention over the recent few years~\cite{dias}-\cite{flach5}. On this ground, significant credit goes to the list of experimental realizations of these geometry sensible \textit{compact localized state} (CLS)~\cite{guzman}-\cite{ram} in a wide class of systems. For this challenging advancement in the fabrication technique, the topic has now extended its relevance beyond the condensed matter physics community - eventually it becomes a general field of discussion in optical~\cite{nolte,real,seba}, acoustic, phononic as well as polaritonic~\cite{amo} systems. The analogous macroscopic degeneracy associated with the FB physics lies at the heart of several novel physical phenomena which covers 
the formation of
fractional quantum hall states~\cite{kapit}-\cite{ando} as well as
 inverse Anderson Localization (AL) transition~\cite{goda}. Synthetic magnetic perturbation in optical networks~\cite{yufan} enables
 the destructive quantum interference resulting in a non-coherent eigenstate (CLS)~\cite{bp2}-\cite{jin}.

Destructive nature of wave interference resulted from the network due to the local geometric phase cancellation invites quantum confinement of wave function within the \textit{characteristic trapping island}. Quantum transport properties is drastically affected for particles
 residing in a flat band. Hence the superposition of flat band modes
 exhibits non-diffrative dynamics of wave packet. Because of divergent effective mass tensor these states do not contribute to overall transmission leading to a singularity in the density of states profile for those non-resonant modes.
There are several crucial factors hindering this morphology induced spatial
imprisonment of excitation including the flatness of the band,
finite lifetime of incoming particles and inter-particle interactions.

Recent study on ultracold fermionic atoms patterned into finite two
dimensional optical lattices of a kagom\'{e} geometry has incorporated renewed interest in this field~\cite{chern}. 
Advancement
in controlled growth of artificial, tailor-made lattices having atoms trapped in optical lattices with the
complications of even a kagom\'{e} structure~\cite{guzman,tamura} has contributed towards increasing interest in these
deceptively simple looking systems. In the context of optics they are very much important as they lead to the long-distance advancement without any kind of distortion of shapes
based on combinations of these flat modes. Actually all these bound states rely on a crucial wave interference condition. 
These contexts
have been demonstrated and analyzed in photonic lattices~\cite{mati,real,seba,mati2}, graphene~\cite{mele,geim}, superconductors~\cite{deng,kohno}, fractional quantum Hall systems~\cite{wen,mudry,gu}, and exciton-polariton condensates~\cite{amo}.
A fascinating feature of non-dispersive systems is the
association of flat band modes to compact localized states
(CLS) which reside on a finite cluster of the lattice and
strictly die out elsewhere. The description of CLS
has been experimentally cited in different flat band systems like frustrated magnets, and photonic crystal
waveguides~\cite{seba2,denz,kle}.
There are a number of recent studies~\cite{flach1,flach4,mielke2} that have discussed for flat band
generating algorithms for giving a general prescription to identify the non-interacting Hamiltonian that supports CLS.

In this communication we study the existence of the \textit{self-localized} modes and discuss the tunability aspect of those non-resonant modes for 3-simplex fractal~\cite{acfrac} geometry. Any fractal entity is readily different from the homogeneous Euclidean objects with respect to its peculiar dimensionality. An alternate demonstration of dimensionality is already mentioned by Hausdorff,
 named as ``Hausdorff Dimension'' where the log-log plot
of two descriptors is termed as dimension~\cite{vince}. Also fractal geometries~\cite{bp1,chacko} have inherent \textit{scale invariance} that can help us to find an analytical prescription to obtain countably infinity of such localized modes and the precise parametric control over those states.
Before going to the specific analysis part  we should expense few words about the motivational aspect details. It is needless to say that these fractal lattices have gained considerable momentum in the condensed matter community because of several application oriented scenario starting from stretchable electronics~\cite{fan} 
to transistors~\cite{ieee}, hydrogen storage~\cite{ruiz}, antennas~\cite{nicola,cohen}, medical imaging~\cite{nz} 
\textcolor{red}{and optimization model for stroke subtypes~\cite{yeliz}}
etc. Moreover, it is also reported that fractals may produce Hofstadter butterfly~\cite{hof}-\cite{smith} and they are also 
suitable prototype example for
showing fractional quantum
Hall states~\cite{pan}, energy level splitting described via
fractal dimensions~\cite{auto} and non-trivial quantum conductivity~\cite{ac,veen}.
 Also, fractals are noticed in self assembled polymers~\cite{bra}.
Although, fractal geometry does not have only theoretical point of interest, rather it has also been exploited experimentally to analyze the magneto-resistance~\cite{gold1}, a distinct crossover of superconducting phase boundary~\cite{gold2}. It has also been used to study the dimensional
crossover and magneto-inductance measurement in Josephson junction arrays of periodically repeated
 Sierpinski gasket fractal network~\cite{lee}.
Unlike the case of ideal deterministic fractal, experimentally fabricated systems cite the scale invariance solely at relatively shorter wavelength fluctuation limit, and it is seen that the
degree of accuracy of the results is highly sensitive to the fractal or
the homogeneous regimes which are being probed. This provides
enough motivation to us to undertake a fractal
type of lattice as our network of interest, but definitely with a slightly different objective.

The basic physical difference between the work related to Sierpinski gasket fractal~\cite{atan}
(reported from our group) and the present 
$3$-simplex fractal lies in their spectral property. Sierpinski gasket fractal is seen to exhibit a set of hierarchical distribution of flat bands but in presence of external magnetic flux, we observed the clear disappearance of such bound states. The spectrum in presence of flux generates resonant bands populated by extended eigen functions. Whereas, in case of simplex fractal structure (present one) the analytical prediction shows that the \textit{self-localized} states no longer vanish but an interesting tunability over the position of such states in respect of application of magnetic flux is demonstrated here. One can continuously modulate the flux, which is an external agency, to control those non-propagating modes. Our aim is to observe the dramatic effect of external parameter on the spectral density of states.

\textcolor{red}{In this communication we have demonstrated an analytical scheme for a deterministic fractal entity to discern the localized states that are non-dispersive in nature. 
Controlled engineering of such states within the tight-binding framework has been discussed elaborately.
It is needless to say that the scheme is in general applicable to several quasi-one dimensional model geometries. The proposition is exact and is consistent with the supportive numerical calculation.}
Here we observe that the position of the flat band may be tampered at will via selective parametric choice and the extent of localization can be manipulated using the self similar pattern of the structure. That means selectively we can block a particular wave train of definite energy using symmetry partitioning and related interference effect and one can easily predict the area of localization from the scale of length used. Real space renormalization group method (RSRG) have been employed to see the multifractal distribution of flat band modes. Since the fractal does not possess any sort of translational ordering, only straightforward diagonalization of the Hamiltonian may not provide the complete spectral information related to FB modes. There is probability that few modes may be slipped away from the original spectrum (in the thermodynamic limit) when one goes for the change in scale of length due to the Cantor-set~\cite{ali} like fragmented spectrum. The method followed here could help in this regard and provide an analytically exact description of FB states for such quasi-one dimensional 
fractal network.
The impact of applying uniform magnetic perturbation is also discussed in terms of the band dispersion. The non-trivial position sensitivity on external parameter makes the discussion challenging to the experimentalists.
\textcolor{red}{Actually, all these analytical results will have physical implication,  specially   when   people   go   for   direct experimental   visualization   of   localization   and other related properties   in low dimensional lattices, and hence this discussion may inspire further experiments on grafted lattices   in   low   dimensions. In view of this, the analogous optical context has been discussed in Sec.~\ref{wg} along with the proposition of single mode wave guide model.}
The equivalent monomode wave guide model is proposed to establish the equivalence of the photonic localization with the same spirit as that in the electronic case.

We thus summarize the complete analysis. In Sec.~\ref{model}, we first describe the model system and the Hamiltonian. Then Sec.~\ref{cls} discusses the basic analytical scheme to obtain the whole hierarchy of the \textit{self-localized} states and the idea of expansion of the extent of \textit{characteristic island}. Sec.~\ref{land} contains usual computation of density of states describing the allowed eigenspectrum. along with
 the discussion of band dispersion. After that in Sec.~\ref{fbflux} we have demonstrated the impact of uniform magnetic flux on the flat band modes. Sec.~\ref{exp} discusses the experimental possibilities, Sec.~\ref{wg}
 demonstrates an equivalent optical wave guide model and finally in Sec.~\ref{conc} we conclude our results.

\section{Model system and Hamiltonian}
\label{model}

\begin{figure}[ht]
\centering
\includegraphics[width=0.9\columnwidth]{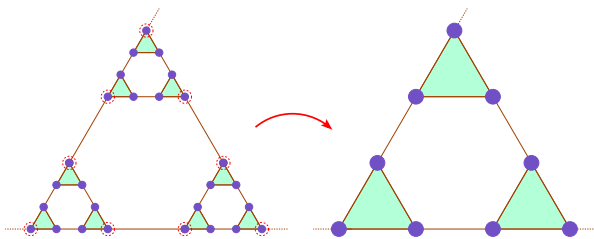}
\caption{(Color online) Schematic representation of the part of an infinite hierarchical 
$3$-simplex fractal geometry
(\textit{left}) and renormalized version of the original network (\textit{right}). 
The sites marked by red dotted circle
are left after one step renormalization.}
\label{lattice}
\end{figure}

We start our demonstration by
referring the Fig.~\ref{lattice} where a finite segment of a geometrically frustrated 
$3$-simplex 
fractal network and its growth sequel are shown pictorially. The quasi-one dimensional hierarchical geometry has
 an inherent self-similar pattern and that leads to an exotic spectral canvas. In the tight-binding language, the overlap parameter between the nearest neighboring atomic sites only is considered in this model. One can easily describe the entire system by the standard Hamiltonian~\cite{acfrac,atan} written in the Wannier basis, viz., 
\begin{equation}
H=\sum_{n} c_{n}^{\dagger} \epsilon_{n} c_{n} + \sum_{\langle nm \rangle} [c_{n}^{\dagger} t_{nm} c_{m} + h.c.]
\label{hamilton}
\end{equation}
The potentials of all the atomic sites are considered to be identical (equal to $\epsilon$) but the hopping integrals are different. The connectivity along the arm of each elemental triangular plaquette is taken as $t$ and one such plaquette is coupled with the other plaquette by the overlap integral $\tau = x t$; $x$ being the coupling parameter. This coupling factor brings the off-diagonal anisotropy. However,
our main objective or point of interest is to see the topological aspects of spectral landscape with the variation of inter-plaquette coupling strength $x$. It is needless to say that the results are no longer sensitive to the numerical values of the parameters of the Hamiltonian. Since we will concentrate on the localization issue induced by the description of the network, we assume, without any loss of generality $\epsilon=0$, $t=1$ for numerical calculation. The tight-binding difference equation~\cite{boy}
 (an alternate form of Schr\"{o}dinger's equation), viz., 
\begin{equation}
(E-\epsilon_{j}) \psi_{j} = \sum_{k} t_{jk} \psi_{k}
\label{diff}
\end{equation}
allows us to analyze the tunability of trapping of electronic wave functions with respect to the modulation of anisotropy parameter $x$.

Following the real space renormalization group (RSRG) scheme, an appropriate subset of the entire lattice can be easily \textit{decimated out} in terms of the amplitudes of the `surviving' nodes to generate the scaled version of the fractal and this can be done by the use of difference equation. In this procedure the parameters of the Hamiltonian may be \textit{renormalized} as follows,
\begin{eqnarray}
\epsilon(n+1)&=&\epsilon(n) + \frac{2 t^2(n) \mathcal{D}_2(n)}{\mathcal{D}_3(n)} \nonumber \\
t(n+1)&=&\frac{t^2(n) T(n) \left[t(n)+\mathcal{D}_1(n)\right]}{\mathcal{D}_3(n)} \nonumber \\
T(n+1)&=&T(n)
\label{rg}
\end{eqnarray}
where $\mathcal{D}_1(n) = [E-\epsilon(n) - T(n)]$, $\mathcal{D}_2(n) = \mathcal{D}_1(n) [E-\epsilon(n)] - t^2(n)$ and 
$\mathcal{D}_3(n) = \mathcal{D}_2(n) [E-\epsilon(n)-t(n)]- T^2(n) \mathcal{D}_1(n)$.
Here
$n \geq 0$ denotes the stage of renormalization.
All the recursion relations in Eq.~\eqref{rg} can now be exploited for explicit analytical 
construction and demonstration of a set
of compact localized states in such deterministic fractal geometry. These equations will also help
us to find out the expected hierarchical distribution of compact localized modes, if any.

\section{Analytical construction of flat band state}
\label{cls}
\subsection{The basic scheme}

One can easily check that by employing the difference equation, if we set 
\begin{equation}
E = \epsilon-t (x+1)
\label{eigen}
\end{equation}
a consistent solution can be achieved for which the non-vanishing amplitudes are concentrated 
\begin{figure}[ht]
\centering
\includegraphics[width=0.9\columnwidth]{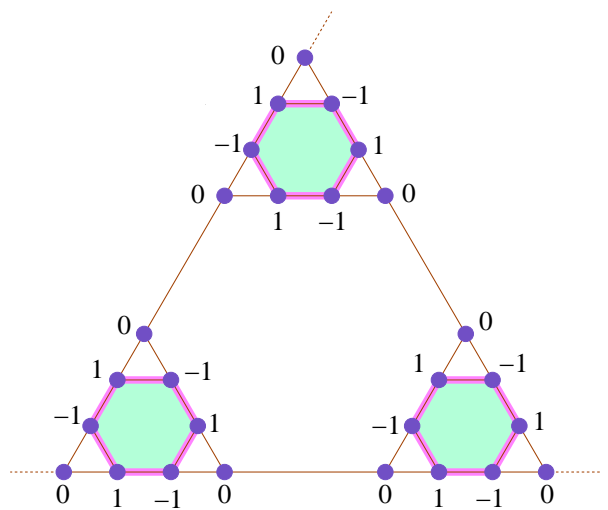}
\caption{(Color online) Wave function amplitude configuration for the compact localized state at $E=\epsilon-t(x+1)$ is shown pictorially.
Each of the shaded hexagonal clusters of atomic sites describes the \textit{characteristic trapping island}
and the connecting nodes with zero amplitudes makes the kinetic information quenched within the 
finite size clusters.}
\label{ampli}
\end{figure}
along the peripheral nodes of the
 \textit{characteristic trapping cell} (hexagonal plaquette here). The connecting node in between any two such clusters has zero wave function amplitude as shown in the Fig.~\ref{ampli}. By making the amplitude vanish at the selective quantum dot locations, it is possible to confine the mobility of the injected wave train (corresponding to the above-mentioned particular value of energy) within a cluster of atomic sites of finite area.
The phase decoherence makes the \textit{prisoner} trapped inside the \textit{characteristic island}.
 Since the dynamics of the wave packet gets locked in space, the analytical construction of those 
\textit{self-localized} modes resembles the essence of a molecular state~\cite{egg}. It is needless to say that the eigenstate is spatially localized due to the formation of physical barrier constituting the sites with zero wave function amplitudes. The destructive quantum interference leads to divergent effective mass tensor and the immobility of the particle. Several clusters are distributed throughout the infinite lattice and one such island is effectively decoupled from the another one making the excitation trapped inside it.
The wave function associated with this eigenmode does not have any overlap with
that described on the neighboring hexagonal plaquette.

\subsection{Self similarity: Flat band hierarchy}
\label{roots}
The underlying fractal geometry bears a self-similar growth sequence. 
\begin{figure}[ht]
\centering
\includegraphics[width=0.7\columnwidth]{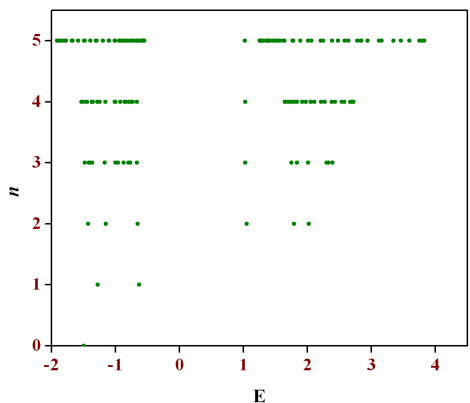}
\caption{(Color online) Hierarchical distribution of localized modes with the index of renormalization
(equivalently the scale of length). The roots form a countably infinite set in the thermodynamic limit.}
\label{fbroot}
\end{figure}
This scale invariance of the structure can be used to construct self trapped 
wave functions at any desirable step of hierarchy. The 
analogous localization area of such bound states can be adjusted at will by a justified choice of RSRG index. 
This is because of the fact that the corresponding eigenvalues can be computed using the following eigenvalue equation for any arbitrary value of $n$, viz.,
\begin{equation}
E = \epsilon(n)-(x+1) t(n) 
\label{clsroot}
\end{equation}
\begin{figure*}[ht]
\centering
(a)\includegraphics[width=0.6\columnwidth]{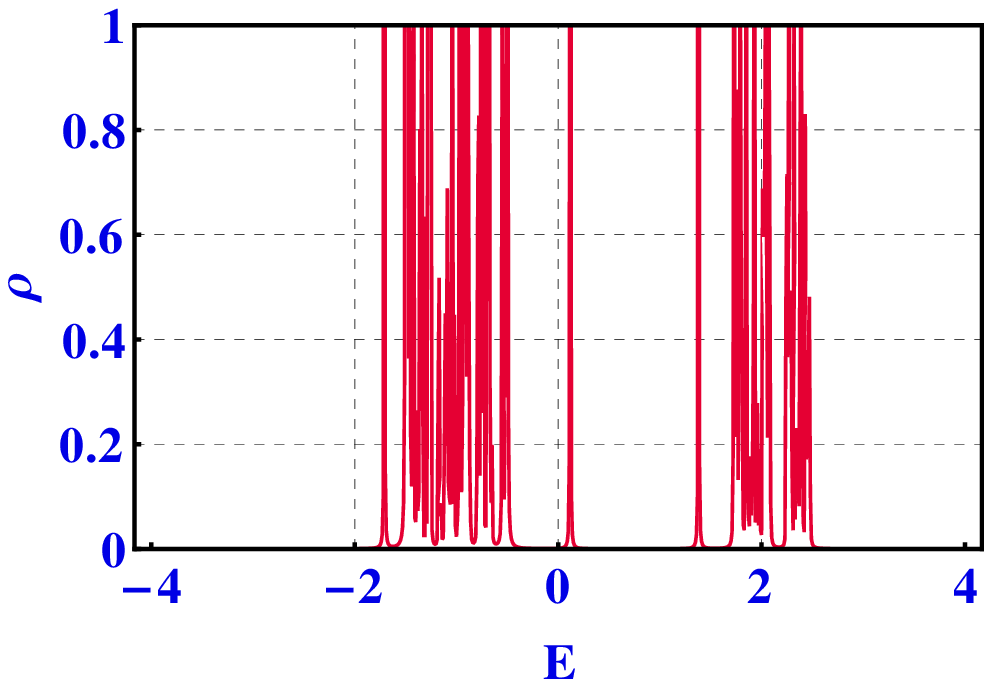}
(b)\includegraphics[width=0.6\columnwidth]{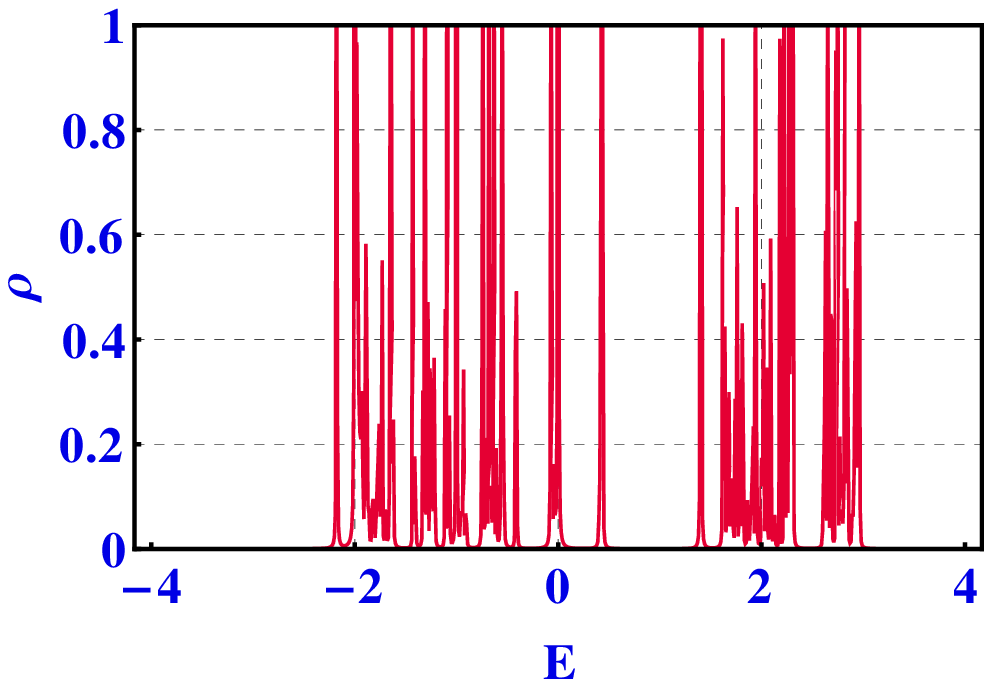}
(c)\includegraphics[width=0.6\columnwidth]{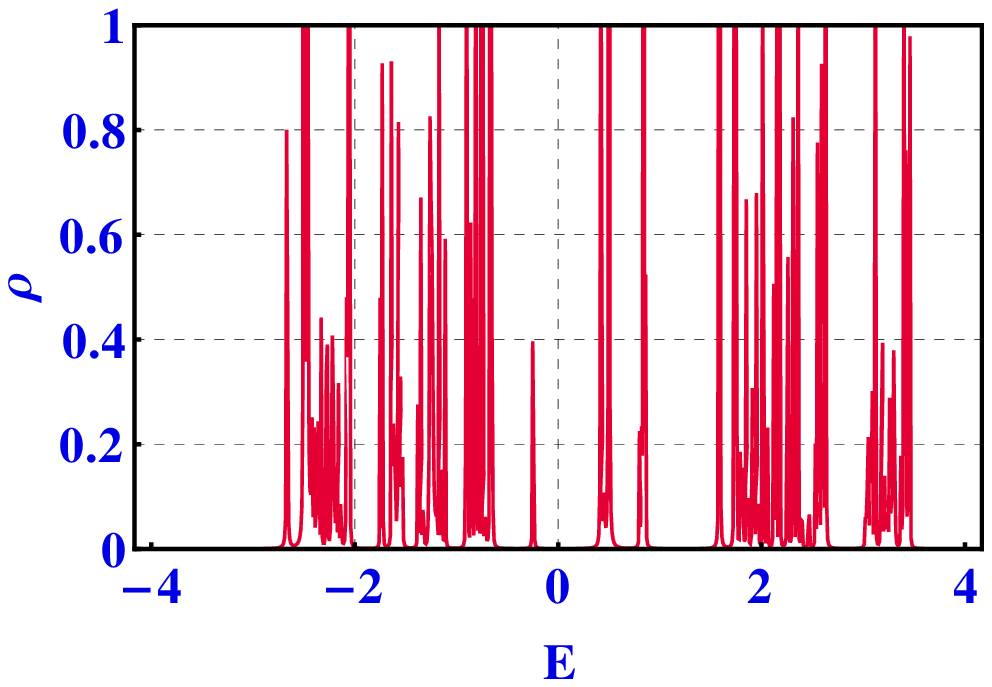}
\caption{(Color online) Plot of density of eigenstates $\rho (E)$ against the energy $E$ of the electron for 
(a) $x=0.5$, (b) $x=1.0$ and (c) $x=1.5$ respectively. The anisotropic parameter $x$ sets the range of
the band extent.The flat bands correspond to sharp spikes at (a) $E=-1.5$, (b) $E=-2$ and (c) $E=-2.5$
respectively. The position of the \textit{self-localized} state is sensitive to the 
anisotropy index $x$. We have set $\epsilon = 0$ and $t=1$ for computational analysis part.}
\label{dos-x}
\end{figure*}
From the above equation we see that the
 localized modes will be an obvious function of the anisotropy parameter $x$. 
Hence the position of these non-coherent bound modes is sensitive on the choice of
the parameter $x$ and thus provides a non-trivial selectivity. This is
expected. 
The energy eigenvalues obtained from the Eq.~\eqref{clsroot} for different stage of hierarchy are plotted in Fig.~\ref{fbroot}. In this demonstration we have particularly set the off-diagonal ansiotropy parameter $x$ as $0.5$. It is seen that as the iteration index increases the number of those self-localized non-resonant modes increases accordingly. The distribution of modes exhibits a typical cantor set-like fragmented spectral sketch which basically reflects the spectral nature of the fractal network. 
We can safely comment that
finally in the thermodynamic limit ($N \rightarrow \infty$) such roots will densely fill up the entire canvas. The sequential clustering of eigenmodes with renormalization index is distinctly cited in Fig.~\ref{fbroot}. It is needless to say that all these states will carry the non-dispersive signature in the overall band spectrum
due to the low group velocity of the wave packet. 
However we will present logical justification in support of this statement in the subsequent discussion.
Now it is to be emphasized that the definite analytical prescription of flat band hierarchy is made possible with the aid of RSRG method only. In this non-translationally invariant lattice, the scheme therefore provides an analytically exact description of at least a subsection of the entire spectrum. Here is the strength of our analytical attempt.
Another interesting scenario related to the nature of localization can be demonstrated as follows.

\subsection{Extension of quantum prison: Staggered localization}

For any energy eigenvalue obtained from any $n$-th level of hierarchy of the underlying fractal geometry, say,
 $n=l$, the hopping integral remains non-vanishing up to that particular $l$-th stage and starts 
decaying beyond it ($n > l$). This signifies that the associated electronic wave functions between the nearest neighboring atomic sites have the non-vanishing overlap at that length scale, When mapped back on to the original fractal network, it definitely amounts to a non-zero overlap of the wave function between the sites much beyond the nearest neighbors (for $n > 1$). That is, the localization of wave packet is based on a cluster of quantum dots, not pinned at any particular node. Such clusters are distributed throughout the geometry, but effectively disconnected from each other, Therefore, the confinement of wave train is \textit{delayed in space}. The effect is named as \textit{staggered localization} 
leading to the creation of a whole set of compact localized states. The spatial extent of the \textit{characteristic unit cell} depends only on the 
iteration index. It is seen that the localization area grows hierarchically with the sequential increment of $n$.

\section{Spectral landscape}
\label{land}
\subsection{Computation of density of states}

Before making any comment on the dispersionless character of the states constructed so far,
we proceed to acquire a general idea of the overall spectrum that describes the entire landscape of allowed eigenmodes as a function of the energy of the injected projectile. It is  clear from 
the Eq.~\eqref{eigen} that for each value of the flat band energies arising out from 
the equation, for any arbitrary value of the coupling strength, it should give a sharp 
distinct spike in the density of states (DOS) spectrum. This is due to the expected 
singularity coming from the zero mobility of the wave train corresponding to those specific bound states
\textcolor{red}{and is governed by the following equation,
\begin{equation}
\rho \propto \int v_{g}^{-1} dk
\end{equation}}
\textcolor{red}{Due to the destructive wave interference the wave packet corresponding to such self-localized modes gets trapped inside the characteristic trapping prison. Hence the group velocity $v_g$ attains a vanishingly small value ($v_g\rightarrow 0$).
The immediate consequence of immobility is the spectral
divergence} and this turns out to be an important predetermining factor in the anomalous behavior of transport, if any. From the above density of modes plots (Fig.~\ref{dos-x}) this singular nature corresponding to
the FB modes is apparent.

The system is a fractal object that carries an inherent scale-invariance. Complete absence of translational 
periodicity makes the appearance of non-resonant localized eigenstates an obvious phenomenon. 
Hence the fragmented kind of cantor set-like spectral pattern is followed from the self-similarity 
present in the underlying structure. One can follow the standard procedure to compute the electronic
 density of states by virtue of the well-used relation~\cite{eco}, viz.,
\begin{equation}
\rho (E) = - \left( \frac{1}{N \pi} \right) Im [Tr G(E)]
\label{dens}
\end{equation}
Here $G(E) = [E-H+i \Delta]^{-1}$ is the usual green's function and $\Delta$ is
 the imaginary part of the energy, comparably small enough, added for the numerical 
 evaluation of DOS. $N$ denotes the system size and `Tr' is the trace of the greens' function. 
 Throughout the numerical analysis, we have set $\epsilon = 0$ and $t = 1$. In Fig.~\ref{dos-x}(a)-(c) 
 we have presented the typical density of states patterns respectively for $x = 0.5$,$1.0$ and $1.5$ 
 as a function of the energy for the quasi-one dimensional hierarchical geometry.
 For small value of the anisotropy parameter $x$ (here $x=0.5$), it cites several distinct spikes
 and very few thin subbands as expected for a fractal entity reflecting its clear distinction from any homogeneous Eucledian objects. The spectrum consists of a number of localized modes distinctly separated by gaps (disallowed modes). For any of the energy eigenvalue
 among any of the discrete spikes, the hopping integral starts diminishing after finite number of loops and this confirms the localization. 
The dynamics of the electrons has
an intricate relationship with the geometric configuration of the underlying fractal lattice. 
The hopping is the off
diagonal element of the Hamiltonian between the Wannier orbitals at neighboring sites, its vanishing clearly demands
that the corresponding electronic wave function cannot extend beyond a certain cluster of minimal dimension. In other words, there is no
overlap between the wave functions beyond a plaquette of certain size. Thus, the amplitudes of the wave function get trapped in
local islands, the islands being distributed and decoupled from each
other all over the fractal geometry. 
 With the gradual increase in the strength $x$, we observe the formation of few mini-patches of allowed energies. However, it retains its multifractal character of the spectrum with the existence of non-resonant states.
 
 The 
features of flat band modes can be easily characterized from the portrait of 
density of states (DOS) in the tight-binding prescription. For each of the FB eigenmodes, 
the non-zero wave function amplitudes are distributed over a finite size atomic cluster and 
those trapping cells accommodating the non-trivial distribution of wave functions get isolated from 
each other beyond a certain scale. The corresponding hopping integral shows decaying behavior with 
the progress 
of iteration loops (scale of length) and this confirms the localization of the associated 
wave packet corresponding to such \textit{self-localized} states.

\subsection{Eigenspectrum as a function of anisotropy parameter}

To complete the discussion we now examine the effect of anisotropy index $x$. For this, we have presented an 
\begin{figure}[ht]
\centering
\includegraphics[width=0.7\columnwidth]{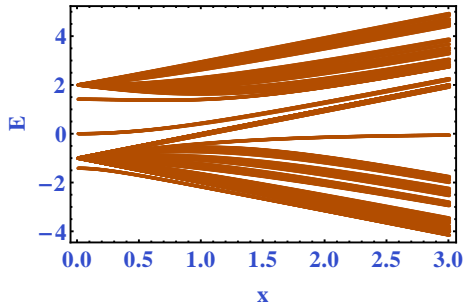}
\caption{(Color online) Allowed eigenspectrum as a function of anisotropy parameter $x$ for a finite size of the 
fractal system.}
\label{exspec}
\end{figure}
allowed eigenvalue distrbution of a $5$-th generation of fractal as a function of $x$ in the Fig.~\ref{exspec}. With the modulation of the parameter $x$, the overlap integral in between any two basic triangular plaquettes changes accordingly. This has an obvious impact on the spectral distribution as seen in the Fig.~\ref{exspec}. The gradual enhancement of anisotropy parameter leads to 
the formation of several allowed minibands and few
selective band overlap. Bandgap is also visible showing the window of disallowed modes. The density of allowed modes is seen to increase with $x$. This is expected because increase in $x$ resonates the inter-plaquette connectivity. We can safely comment that the density of states is larger in the regime $t \leq x \leq 3t$ and this tradition is expected to continue beyond this range ($x \ge 3t$) exhibiting more number of band merging. This 
non-trivial variation with $x$ provides a tunability of those 
\textit{self-localized} states that could be interesting from the experimental point of view.

\subsection{Band dispersion}
The underlying geometry does not have any translational ordering and hence the deduction of energy-momentum
\begin{figure}[ht]
\centering
\includegraphics[width=0.85\columnwidth]{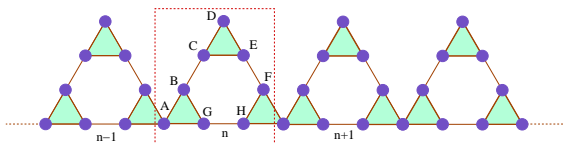}
\caption{(Color online) Schematic representation of the periodic array of fractal geometry of first generation. }
\label{array}
\end{figure}
 relationship in usual tight-binding language is no longer possible. This prompts us to undertake an indirect but effective way to discern the non-dispersive flavor of the whole hierarchy of those self-localized states.
The motivation behind this analysis is that the compact localized states constructed in the earlier section display a typical self-localization induced by the nature of the resultant wave interference. The trapping of amplitudes inside the \textit{allowed zone} makes the particle `super-heavy' (infinite effective mass) such that it loses its mobility. This immediately offers vanishing band curvature corresponding to the particular eigenmode and hence carries a non-dispersive signal. 
\begin{figure*}[ht]
\centering
(a)\includegraphics[width=0.6\columnwidth]{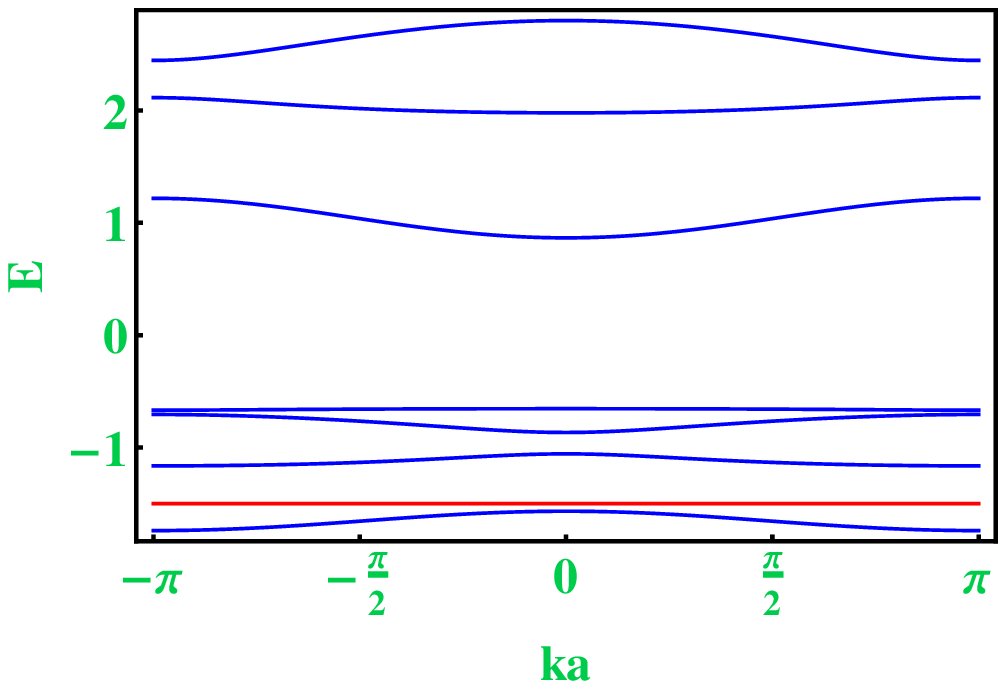}
(b)\includegraphics[width=0.6\columnwidth]{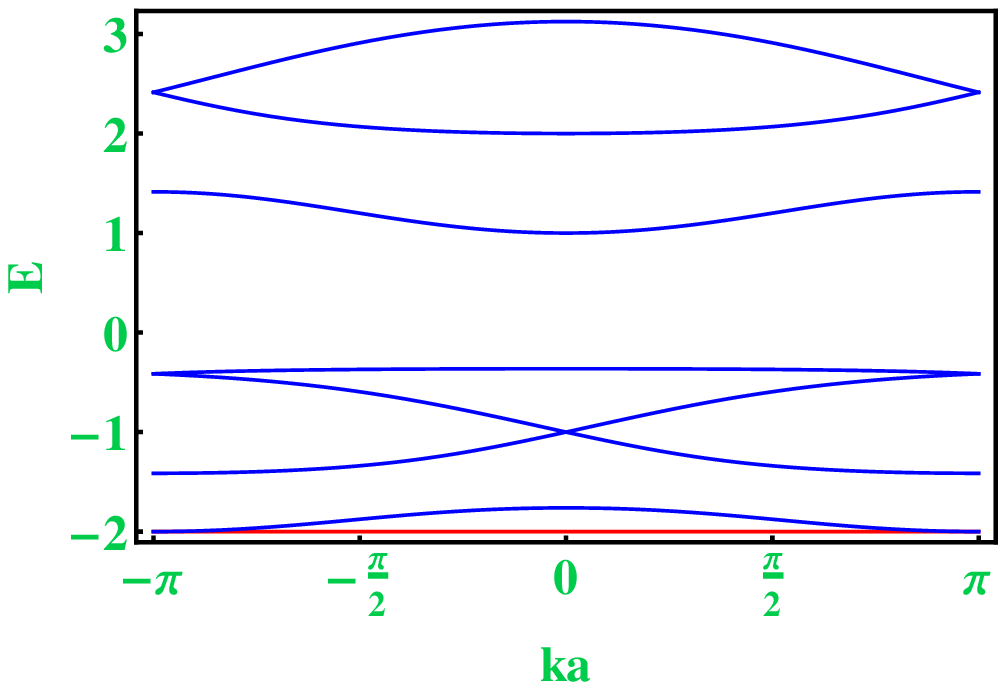}
(c)\includegraphics[width=0.6\columnwidth]{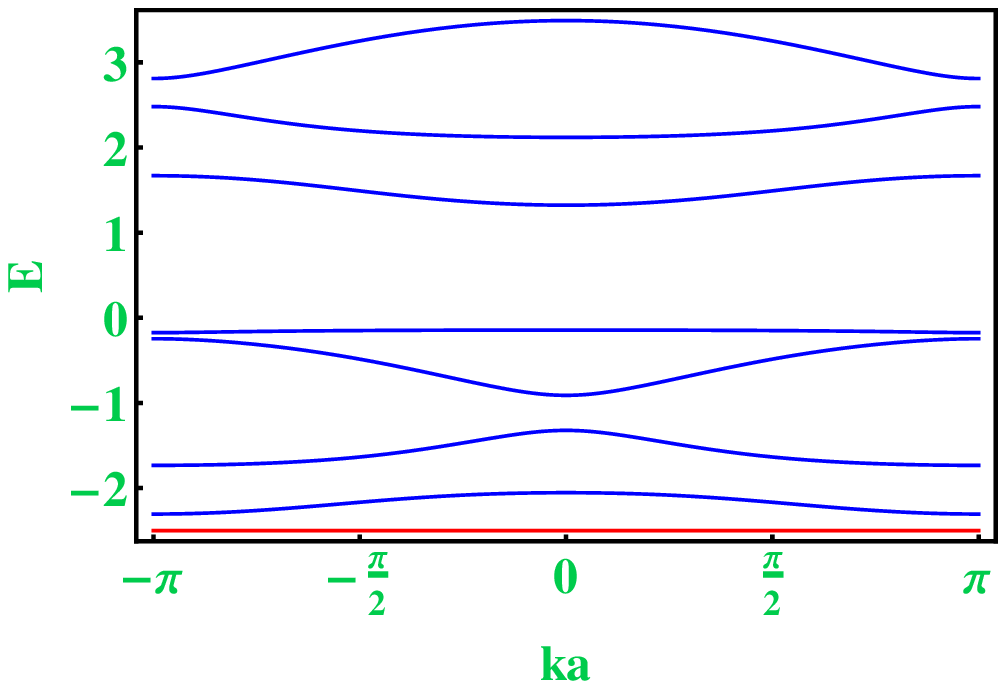}
\caption{(Color online) Energy-momentum relationship plotted for a quasi-one dimensional array of 
simplex fractal geometry of first generation for (a) $x=0.5$, (b) $x=1.0$ and (c) $x=1.5$ respectively. The red line in each 
case denotes the corresponding non-dispersive flat band satisfying the eigenvalue equation (Eq.~\eqref{eigen}).
The number of bands will increase sequentially with the level of hierarchy.}
\label{disp1}
\end{figure*}

We choose a periodic array as depicted in Fig.~\ref{array} 
by considering a finite generation of simplex fractal as unit cell. For this array, we can recast the Hamiltonian in the momentum space~\cite{bp1}, viz.,
\begin{equation}
H = \sum_{k} \psi^{\dagger}_{k} \mathcal{H} (\bm{k}) \psi_{k}
\label{ktransform}
\end{equation}
This typical construction creates a top vertex with coordination number equal to two. Though it was not present in the infinite geometry where we cannot see the ends, but when we will upgrade the generation of the fractal, in the true thermodynamic limit, this will act as a reasonably good approximant and carry the flavor of the original system. Our ultimate goal is to prove the dispersionless nature of the compact localized modes for this geometrically frustrated fractal entity. Therefore it is expected that in the language of reciprocal space, the momentum insensitive part of the band dispersion should produce the same bound states. By the conversion, Hamiltonian matrix reads as,
\begin{equation}
\mathcal{H}(\bm{k})=
\left[ \begin{array}{cccccccc}
\epsilon_A & t & 0 & 0 & 0 & te^{-ik} & t & te^{-ik}\\
t & \epsilon_B & xt & 0 & 0 & 0 & t & 0\\
0 & xt & \epsilon_C & t & t & 0 & 0 & 0\\
0 & 0 & t & \epsilon_D & t & 0 & 0 & 0\\
0 & 0 & t & t & \epsilon_E & xt & 0 & 0\\
te^{ik} & 0 & 0 & 0 & xt & \epsilon_F & 0 & t\\
t & t & 0 & 0 & 0 & 0 & \epsilon_G & xt\\
te^{ik} & 0 & 0 & 0 & 0 & t & xt & \epsilon_H
\end{array}
\right]
\end{equation}
We have set identical potentials for all the sites and carried out the dispersion equation.
For this quasi-one dimensional fractal structure we have plotted the band dispersion relation by virtue 
of the above matrix (in reciprocal space) for different values of $x$ as cited in the Fig.~\ref{disp1}. Here we find that the band dispersion curves readily confirms the dispersionless behavior of the analytically constructed self localized states. The red lines in all the cases show the flat bands satisfying the analytical prescription (Eq.~\eqref{eigen}). Even with
a modest unit cell combination for the array, the staggered localized states corresponding to
\begin{figure}[ht]
\centering
(a)\includegraphics[width=0.4\columnwidth]{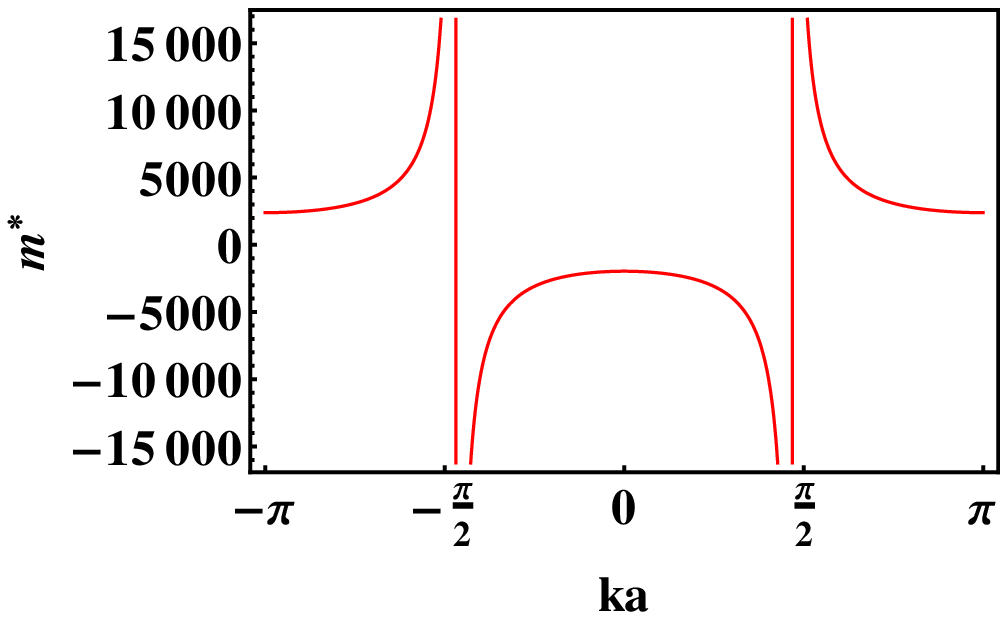}
(b)\includegraphics[width=0.4\columnwidth]{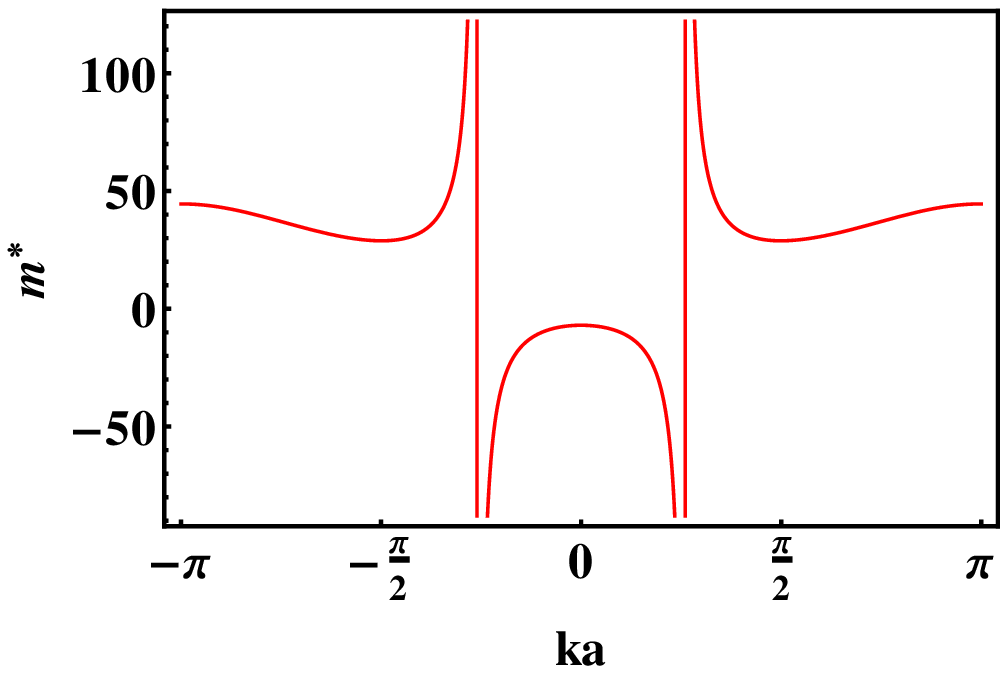}\\
(c)\includegraphics[width=0.4\columnwidth]{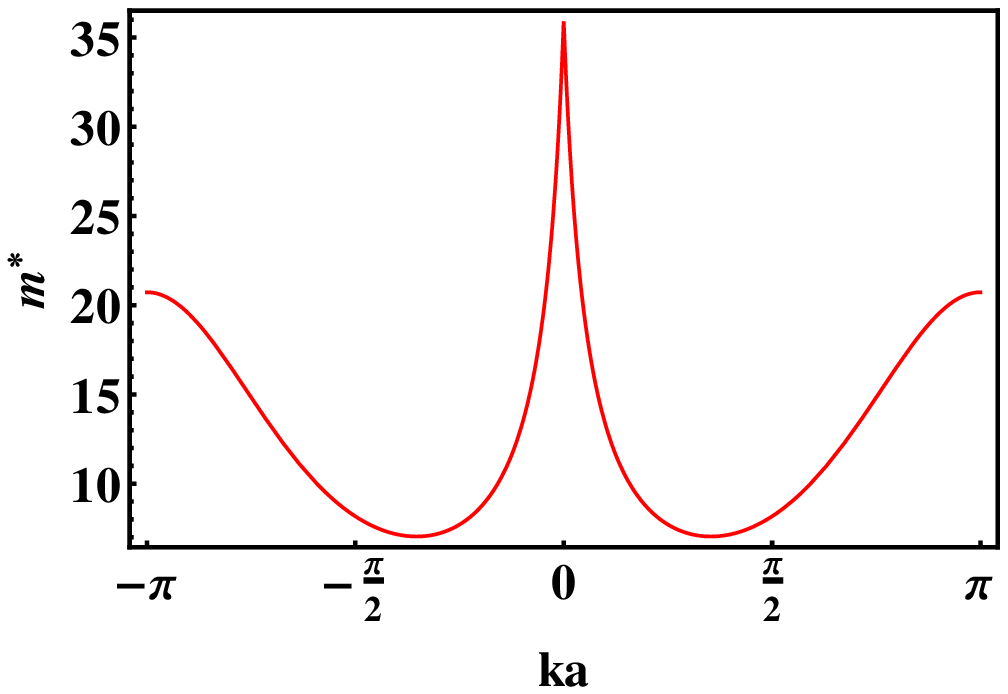}
(d)\includegraphics[width=0.4\columnwidth]{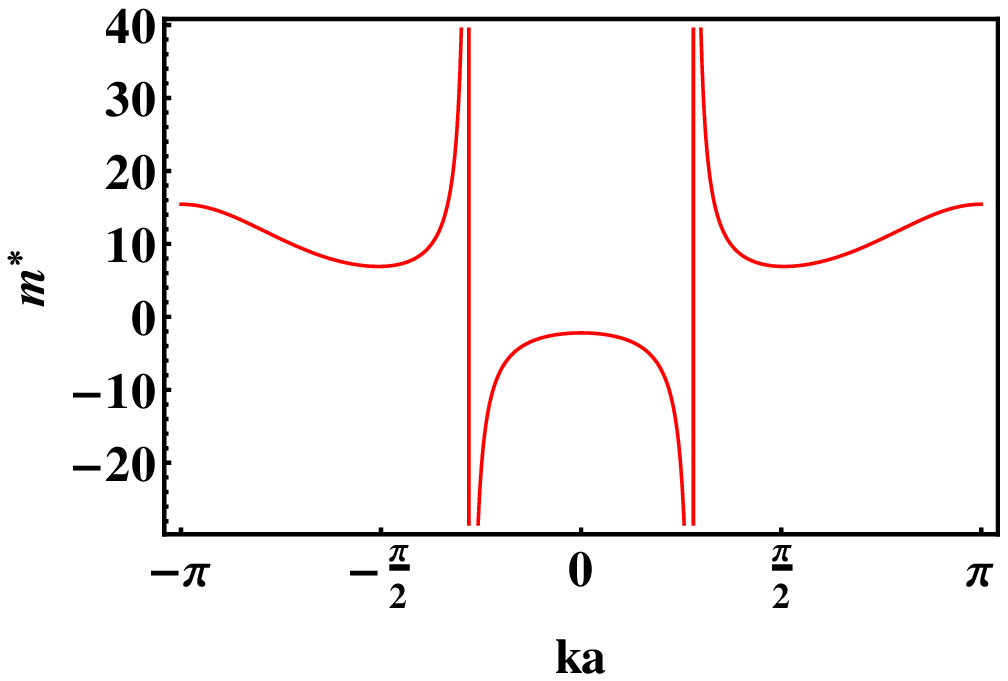}
\caption{(Color online) Effective mass variation for the fractal geometry for
(a) $x=0.05$, (b) $x=0.5$, (c) $x=1$ and (d) $x=2$ respectively.}
\label{efmass}
\end{figure}
the truly infinite system are showing up. The plots exhibit other dispersive bands and quasi-dispersive (bands with extremely low but non-zero curvature). For the later case, the electrons carrying those particular energies have significantly low mobility. Also it is observed that in absence of asymmetry ($x=1$), there is a band crossing which is seen to be missing in the next plot ($x=1.5$) and thus it immediately creates an energy gap in between the two bands with the increment of anisotropy index. It is needless to say that the number of bands
(both dispersive and non-dispersive) will increase sequentially with the 
increase in hierarchy and all those modes will densely fill up the band structure. 
Also all the localized modes will form a set of hierarchy 
comprising of countably infinite number of such modes in the thermodynamic limit.
This message is
clear from the Fig.~\ref{fbroot} and the associated discussion. 

\subsection{Parameter selective control of effective mass}

From the band dispersion plot we see the existence of some dispersive states. Of course the position is highly
sensitive on the value of anisotropy parameter $x$. Hence using that factor one may tune the group velocity of the wave packet as well as the band curvature (equivalently effective mass tensor of the particle). Thus parameter induced engineering of effective mass essentially invites the precise control over mobility of the incoming projectile. Fig.~\ref{efmass} represents the variation of effective mass for different strength of the coupling parameter for one such resonant mode. From the plots we see that for low enough value of $x$, the effective mass of the particle shows extremely large magnitude and hence the group velocity becomes considerably small. This leads to localization of wave train. With the gradual increment magnitude of effective mass tends to reach some finite values, but still the variation is oscillating in nature between both positive and negative ranges. When the off-diagonal anisotropy is absent ($x=1$) oscillation is seen to be restricted within some positive finite value and again for moderate or strong enough anisotropy index ($x > 1$), again the sign flipping scenario comes back within finite range of values. Also we observe several diverging behavior of mass for few k-points. Another interesting fact is the \textit{re-entrant} fashion from positive to negative regime through those singular points. The parameter selective sharp cross-over between the hole like and electron like character of the particle is the main physical implication of this discussion. One may tune the electron-lattice interaction with the choice of system parameter.

\section{Impact of perturbation on flat band}
\label{fbflux}
\subsection{Flux sensitive Aharonov-Bohm caging}

In the previous section, we have discussed the effect of off-diagonal anisotropy on the band dispersion. Now we will switch off this anisotropy effect and make the hopping uniform everywhere. The central aim of the subsequent discussion is to study the sustainability of the FB in presence of homogeneous magnetic perturbation piercing through any local closed plaquette of the geometry. All the flat band states carry 
macroscopic degeneracy with a basis of eigenstates which are spatially
 localized since the analogous wave function form a standing-wave pattern because of the incoherent wave interference. This extensive degeneracy imports an obvious sensitivity to many kinds of external
perturbation, including disorder (deterministic or random), different types of interactions or external fields. Based on the response to these kind of external parameter, a clear distinction is obtained by Aoki et al. to classify the FB networks~\cite{ando}. Uniform
magnetic field breaks the macroscopic degeneracy; destroys the FB in Mielke's and Tasaki's lattices~\cite{mielke}-\cite{tasaki}
and transforms it into a pattern of Hofstadter butterfly. On the contrary, Lieb or Dice class of networks support chiral FB~\cite{flach1} which do not respond to the external agency of perturbation due to the symmetry protection. Because this sustainability involves ``local symmetry'' of the sub-lattice connection. The later class of superlattices exhibiting symmetry-protected chiral flat dispersionless band was first cited by
Aoki et al.~\cite{shima}. The undisturbed FB state protect their extensive degeneracy, though there may be position shifting with respect to the application of magnetic field. This topic has gained considerable interest since $2012$~\cite{lan,zhu}. Also we should mention that the flux dependent flat band is a subset of a widely
known phenomena, i.e., Aharonov-Bohm caging~\cite{vidal} and has been recently verified experimentally by a group of 
workers~\cite{sebanew}.

In our geometrically frustrated fractal system, the uniform magnetic flux is trapped inside each basic
hexagonal plaquette and this eventually breaks the time reversal symmetry (at least locally) 
of the off-diagonal element of Hamiltonian along the bond. The incorporation of magnetic flux $\Phi$ introduces Peierls' phase factor associated with the overlap parameter, viz., $t \rightarrow t e^{i \Theta}$, 
where $\Theta =  (\pi \Phi/ 3 \Phi_{0})$, $\Phi_0 = hc/e$ is known as fundamental flux quantum. It is to be noted that to see the effect of applying perturbation, we set $x=1$. Therefore the hopping integral is uniform 
magnitude-wise everywhere in lattice. Now we
\begin{figure}[ht]
\centering
\includegraphics[width=0.6\columnwidth]{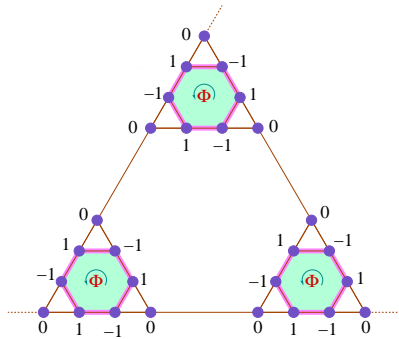}
\caption{(Color online) Amplitude distribution for flux sensible flat band at 
$E = \epsilon -2 t \cos \Theta$. The macroscopic degeneracy associated with the flat bands is
sustained in presence of perturbation. We have set the 
\textit{anisotropy index} $x=1$ to see the flux tunability of flat bands.}
\label{ampli2}
\end{figure}
 switch on the magnetic perturbation $\Phi$. By virtue of the difference equation it is quite trivial to check that if we take $E = \epsilon-2 t \cos \Theta$, then a flux dependent amplitude configuration can be drawn
\begin{figure}[ht]
\centering
(a)\includegraphics[width=0.6\columnwidth]{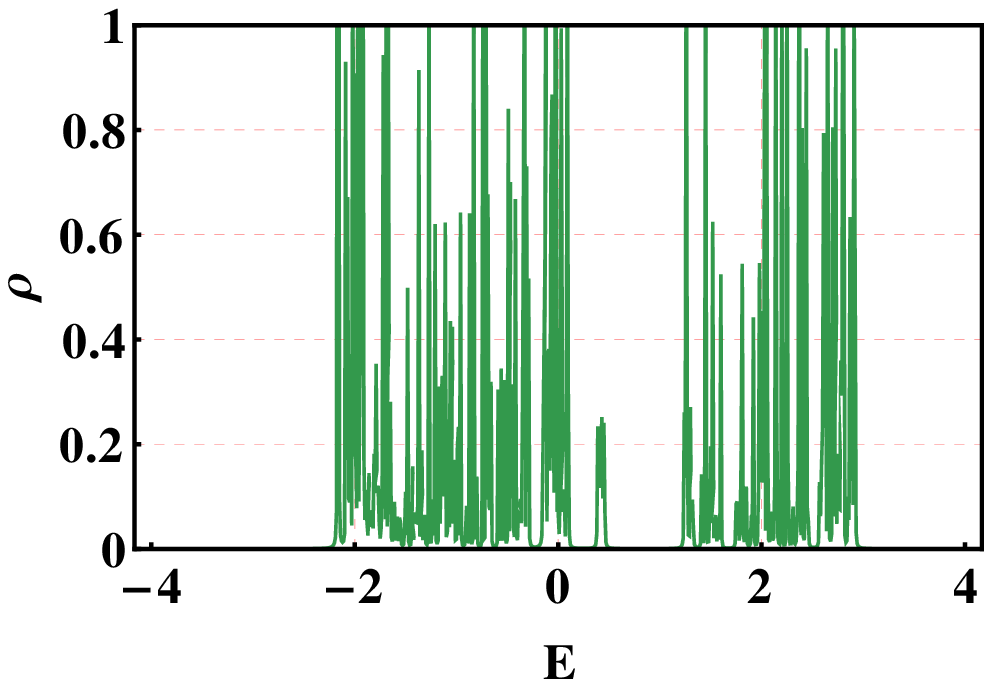}\\
(b)\includegraphics[width=0.6\columnwidth]{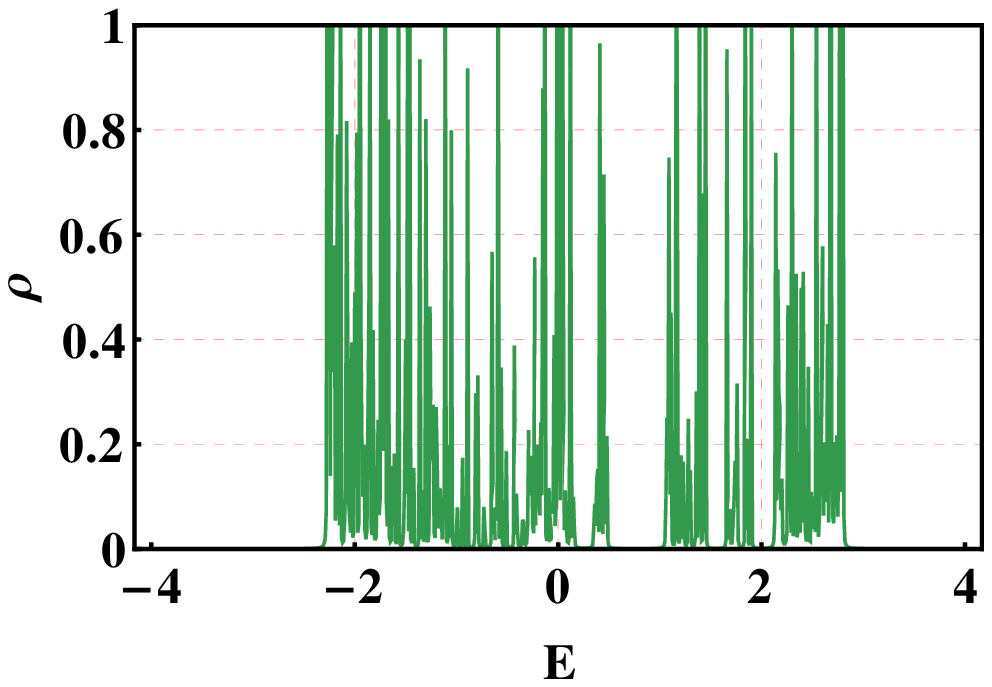}
\caption{(Color online) Variation of density of states $\rho (E)$ as a function of the energy $E$ of 
the incoming electron for  $x=1$. The magnetic flux threading each elemental triangular plaquette is set at (a) $\Phi=\Phi_{0}/4$ and (b) $\Phi=\Phi_{0}/2$ respectively. $\Phi_0 = h c/e$ is the fundamental flux quantum.}
\label{dos-phi}
\end{figure}
 satisfying the difference equation and is shown in Fig.~\ref{ampli2}. The resultant nature of the quantum interference obtained from multiple quantum dots solely determines the sustainability issue of the FB modes. From the above equation it is clear that with the application of homogeneous magnetic flux, the position of the FB is now extremely sensitive to the applied flux. Hence one can tune the external parameter \textit{at will}
  to trap any particular wave packet of definite energy eigenvalue in a selective way. This means for the injected projectiles to be locked via lattice structure the energy must have to satisfy the above eigenvalue equation for a particular value of flux. The position engineering controlled by the periodic modulation of flux makes this discussion interesting from the experimental perspective. Another point is that the above flux involved amplitude distribution scenario is valid for any finite value of flux. Therefore, one can say the extensive degeneracy associated with the FB states is not disturbed anyway by the application of
  flux. In Fig.~\ref{dos-phi} we have shown the density of eigenstates plots for two different values of magnetic flux $\Phi = \Phi_{0}/4$ and $\Phi_{0}/2$ respectively. It is needless to mention that owing to the vanishing mobility of the wave train corresponding to FB mode, the DOS presents a distinct divergent behavior
  for the selective states satisfying the flux-dependent equation. This
  singularity confirms the prediction of flux modulated flat band states. Of course, this singularity is now external agency dependent.

\subsection{Allowed eigenspectrum}
\begin{figure}[ht]
\centering
\includegraphics[width=0.9\columnwidth]{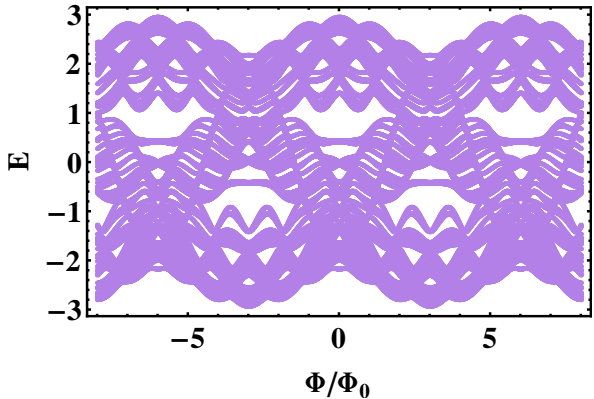}
\caption{(Color online) Flux sensitive allowed eigenspectrum as a function of external perturbation.
The spectrum is flux periodic with periodicity equal to one flux quantum. We have set the 
\textit{anisotropy index} $x=1$ to see the flux tunability of flat bands.
The flux tunable flat band at $E=\epsilon - 2 t \cos \Theta$ is seen to be present.}
\label{spectrum}
\end{figure}

For the sake of completeness we should present a distribution showing the allowed eigenmodes as a function of
 external magnetic flux (Fig.~\ref{spectrum}). This figure also supports the existence of flux tunable flat band.
  This diagram may be considered an analogous dispersion relation because for an electron moving round a closed path, it is well known that the magnetic flux plays the similar physical role as that of the wave vector~\cite{gefen}. The eigenmodes create miniband-like structure as a function of external flux. The pattern presented contains an inherent periodicity equal to six times the fundamental
  flux quantum $\Phi_0$. First of all, the eigenspectrum 
supports the existence of flux sensitive flat band modes as suggested in Fig.~\ref{ampli2}.  
  One can continuously engineer the magnetic flux to adjust the imprisonment of wave train with 
  high selectivity.
  Moreover, there are a number of inter-twined band overlap, and a quite densely packed distribution of allowed modes, forming quasi-continuous flux-periodic $E-\Phi$ band structure. 
  Close observation of this eigenspectrum reveals the
formation of interesting variants of the Hofstadter butterflies~\cite{hof}.
The spectral landscape is a quantum fractal, and encoding the gaps with appropriate topological quantum numbers remains an open problem for such
 deterministic fractals.

\section{Discussion regarding experimental aspects}
\label{exp}

The essence of quantum imprisonment of incoming wave packet by means of lattice geometry and wave interference resulted from the network is no longer a theoretical issue in the present era of advanced nanotechnology and lithography techniques. Thanks to the experimentalists that they have made it possible to visualize the authentic scenario related to the localization of injected messenger in  laboratory with the help of femto second laser-writing procedure and optical induction technique. Consequently, the engineering of localization of wave train is a very common aspect in the different topics of condensed matter physics. Also it is very pertinent in case of classical electromagnetic signal and light as well. The experimental progress started with Yablonovitch~\cite{yab1,yab2} along with the propositions from John~\cite{john}
 and Pendry and MacKinon~\cite{pend} related to the realistic experimental observation of Anderson localization.
 This essentially invites the path-breaking idea of slow light engineering~\cite{baba} that opens up the
 possibility of \textit{spatial compression of light energy}.
This area of research has also enriched by the most recent examples~\cite{seba,seba2,sebanew} of photonic localization~\cite{hu,hu2} in some one or two dimensional geometries. This gives a scope for direct observation of diffraction free FB states. All these challenging experiments have become the milestone in this field of study and it is needless to say that these inspire us to take such model fractal object as our system of interest.
The predictability of the flat band engineering with respect to the anisotropy index and magnetic flux as well,
 throws an achievable challenge to the experimentalists to test the analytical result with proper modulation. 
 This may be possible in accordance with
  the proposition of S. Longhi~\cite{longhi1,longhi2}, where it was shown that the introduction of the concept of flux 
 controlled band engineering may be possible as a synthetic gauge field 
 may be fabricated by changing the propagation constant and this could help in the experimental 
 possibility.

\section{Proposed optical wave guide model}
\label{wg}
Exact analogy between the electronic case and the associated optical scenario within the tight-binding formalism is very useful as well as pertinent aspect.
 The concept was first initiated by Sheng et al.~\cite{alexan,sheng}, where they showed that the propagation of wave train through any quasi-one dimensional monomode wave guide structure has a direct correspondence with the same electronic model with proper initial conditions. 
\begin{figure}[ht]
\centering
\includegraphics[width=0.75\columnwidth]{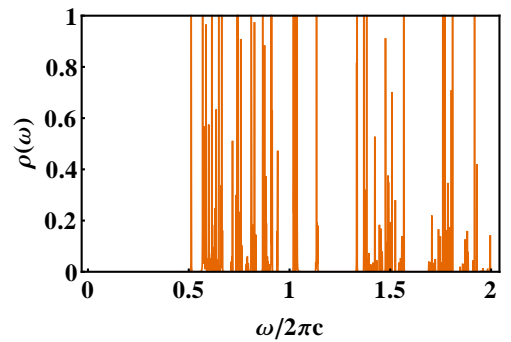}
\caption{(Color online) Representation of density of photonic modes as a function of the frequency of the
incoming wave. Frequency $\omega$ is taken in unit of $2 \pi c/a$, $a$ being the wave guide dimension.}
\label{optdos}
\end{figure}
The analogy of the solution of the wave equation with the difference equation for an electron moving in an identical geometry provides a parallel platform to study the localization issue of classical waves. 
Also, it is needless to say that the
 one-to-one mapping is completely a mathematical artifact and once that is done the recursion relations exploited for RSRG approach
 become insensitive to whether the input appears from a quantum background or a classical one. 
We should point out that the propagation of signal can be adjusted and controlled by appropriate selection of dielectric parameter of the core material.
If we consider a definite dimension of the wave guide segment from the very beginning, then
 it is quite trivial to obtain the frequency of incoming wave packet
 corresponding to the localized modes.
 
Following Sheng et al., one can fabricate a single channel wave guide model formed by the segments
having the same dimensions arranged in a simplex fractal geometry.
Within the tight-binding framework we have plotted (Fig.~\ref{optdos}) the variation of density of optical modes against the frequency $\omega$ of the incoming wave to acquire knowledge about the
 general idea about the allowed eigenmodes
for the analogous fractal wave guide network.
The distribution of optical modes within the frequency regime $0< \omega < 2$ shows several clusters of non-zero values of density of photonic modes. The numerical value of the dielectric constant $\epsilon_r$ is arbitrarily set as $2$ and the anisotropy in the hopping (as discussed in the electronic case) is introduced by incorporating a wave guide segment of same dimension but with different dielectric parameter ($\epsilon_r =3$) in the suitable place.
The spectrum is highly fragmented in character and this is expected for a fractal structure.
The appearance of localized modes triggered by destructive wave interference is clear in the graphical presentation.
Also, the
 density of modes spectrum exhibits few gaps in the frequency range for which the
 wave propagation is not permissible. Hence this proposed model may act as an appropriate candidate for photonic bandgap (PBG)~\cite{yab1,yab2} kind of system.

\section{Summary and outlook}
\label{conc}
We have described the analytical scheme following real space renormalization group technique to compute the hierarchical distribution of dispersionless flat band eigenmodes for geometrically frustrated kind of $3$-simplex fractal lattice. These states are essentially localized over clusters of atomic sites by means of phase cancellation at some connecting nodes induced by destructive wave interefrence. The extent of imprisonment of wave packet may be controlled at will via selective choice of iteration index (or scale of length). The prescription presented here provides an exact set of countably infinite number of bound states as a function of off-diagonal anisotropy parameter. The numerical evaluation of density of eigenstates and band dispersion corroborate our analytical proposition. The impact of applying uniform magnetic perturbation is discussed elaborately. Flux tunable position sensitive macroscopically degenerate flat bands are seen in this fractal geometry. Continuous periodic modulation of external flux over the non-resonant modes eventually invites flux sensitive localization for this type of quasi-one dimensional network. The inherent singularity corresponding to those modes is also seen to be present in the spectral landscape. Hierarchical distribution of flat band modes in the thermodynamic limit is basically compatible with the fragmented nature of eigenspectrum.

\begin{acknowledgments}
The author is thankful to Ms. Amrita Mukherjee for giving helpful suggestions regarding 
computational aspects. The author also gratefully acknowledges the stimulating discussions
related to the part of the results with Ms. Mukherjee.
\end{acknowledgments} 

\end{document}